\pdfoutput=1
\documentclass[acmsmall,screen]{acmart}

\usepackage{booktabs}
\usepackage{tikz}
\usepackage{xcolor}
\usepackage{listings}
\usepackage[utf8]{inputenc}

\definecolor{lstkw}{HTML}{0B5394}   
\definecolor{lststr}{HTML}{8B4513}  
\definecolor{lstcom}{HTML}{6E7681}  

\lstdefinelanguage{ImpHs}{%
  morekeywords={%
    case, of, do, let, in, if, then, else, where,%
    data, type, family, newtype, class, instance, deriving,%
    forall, return, module, import, qualified, as, hiding,%
    Imp.do, Imp.return%
  },%
  sensitive=true,%
  alsoletter={.},%
  morecomment=[l]{--},%
  morecomment=[s]{\{-}{-\}},%
  morestring=[b]"%
}

\lstdefinestyle{ImpStyle}{%
  language=ImpHs,
  basicstyle=\ttfamily\small,
  keywordstyle=\color{lstkw}\bfseries,
  commentstyle=\color{lstcom}\itshape,
  stringstyle=\color{lststr},
  showstringspaces=false,
  columns=fullflexible,
  keepspaces=true,
  xleftmargin=1em,
  aboveskip=\medskipamount,
  belowskip=\medskipamount,
  literate=%
    {<-}{{$\,\leftarrow\,$}}2%
    {->}{{$\,\rightarrow\,$}}2%
    {=>}{{$\,\Rightarrow\,$}}2%
    {§}{{\$}}1
    {λ}{{\textbackslash}}1
}

\lstnewenvironment{code}{\lstset{style=ImpStyle}}{}

\lstnewenvironment{plain}{%
  \lstset{%
    basicstyle=\ttfamily\small,%
    columns=fullflexible,%
    keepspaces=true,%
    xleftmargin=1em,%
    aboveskip=\medskipamount,%
    belowskip=\medskipamount%
  }%
}{}

\newcommand{\hs}[1]{\texttt{#1}}
\newcommand{\true}{{\hs{True}}}
\newcommand{\false}{{\hs{False}}}
\newcommand{\WMC}{\mathsf{WMC}}

\setcopyright{cc}
\setcctype{by}
\acmJournal{PACMPL}
\acmYear{2026} \acmVolume{10} \acmNumber{ICFP} \acmArticle{300}
\acmMonth{8} \acmDOI{10.1145/3828698}
\acmSubmissionID{icfp26main-p87-p}
\received{2026-02-19}
\received[accepted]{2026-05-13}

\begin{document}

\title[Imprecise Probabilistic Programming, Precisely]{Imprecise Probabilistic Programming, Precisely\\ (Functional Pearl)}
\subtitle{Credal Sets via Graded Monads, BDDs, and Semiring-Parametric Inference}

\author{Jack Liell-Cock}
\orcid{0009-0005-7121-8095}
\affiliation{%
  \institution{University of Oxford}
  \city{Oxford}
  \country{United Kingdom}
}
\email{jack.liell-cock@cs.ox.ac.uk}

\author{Sam Staton}
\orcid{0000-0002-7149-3805}
\affiliation{%
  \institution{University of Oxford}
  \city{Oxford}
  \country{United Kingdom}
}
\email{sam.staton@cs.ox.ac.uk}

\begin{abstract}
Imprecise probability generalizes standard probability theory by replacing a single distribution with a convex set of possible distributions.
We show that this generalization requires no change to the standard BDD compilation and weighted model counting pipeline used by discrete probabilistic languages.
An imprecise coin flip is simply a BDD variable whose weight is left free rather than fixed.
We introduce \textsc{Imp}, a Haskell-embedded DSL for imprecise probabilistic programming.
A graded monad, indexed by finite sets of named sources of epistemic uncertainty,
restores the commutativity that the standard convex powerset monad lacks, and GHC's type system enforces this at compile time.
Weighted model counting is parametric in the semiring, so the same compiled BDD supports exact, differentiable, and interval-bounded inference.
\end{abstract}



\maketitle

\section{Introduction}
\label{sec:intro}

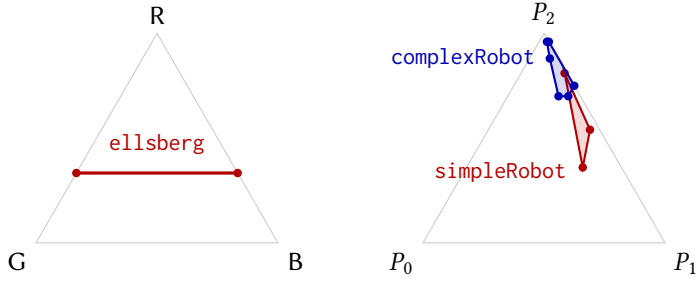
\begin{figure}[t]
\centering
\begin{tikzpicture}[scale=3.2]
  \begin{scope}[shift={(-0.7, 0)}]
    \coordinate (Gv) at (0, 0);
    \coordinate (Bv) at (1, 0);
    \coordinate (Rv) at (0.5, {sqrt(3)/2});
    \draw[gray!40] (Gv) -- (Bv) -- (Rv) -- cycle;
    \node[below left]  at (Gv) {$\mathsf{G}$};
    \node[below right] at (Bv) {$\mathsf{B}$};
    \node[above]       at (Rv) {$\mathsf{R}$};

    \coordinate (E1) at ({1/6}, {sqrt(3)/6});
    \coordinate (E2) at ({5/6}, {sqrt(3)/6});
    \draw[red!70!black, very thick] (E1) -- (E2);
    \fill[red!70!black] (E1) circle (0.5pt);
    \fill[red!70!black] (E2) circle (0.5pt);
    \node[red!70!black, above, font=\small] at (0.5, {sqrt(3)/6 + 0.04}) {\hs{ellsberg}};
  \end{scope}

  \begin{scope}[shift={(0.9, 0)}]
    \coordinate (Rv) at (0, 0);
    \coordinate (Gv) at (1, 0);
    \coordinate (Bv) at (0.5, {sqrt(3)/2});
    \draw[gray!40] (Rv) -- (Gv) -- (Bv) -- cycle;
    \node[below left]  at (Rv) {$P_0$};
    \node[below right] at (Gv) {$P_1$};
    \node[above]       at (Bv) {$P_2$};

    \coordinate (R2a) at (0.66, {0.36*sqrt(3)/2});
    \coordinate (R2b) at (0.69, {0.54*sqrt(3)/2});
    \coordinate (R2c) at (0.585, {0.81*sqrt(3)/2});
    \fill[red!70!black, opacity=0.15] (R2a) -- (R2b) -- (R2c) -- cycle;
    \draw[red!70!black, thick] (R2a) -- (R2b) -- (R2c) -- cycle;
    \fill[red!70!black] (R2a) circle (0.5pt);
    \fill[red!70!black] (R2b) circle (0.5pt);
    \fill[red!70!black] (R2c) circle (0.5pt);
    \node[red!70!black, left, font=\small] at (0.63, {0.36*sqrt(3)/2 - 0.02}) {\hs{simpleRobot}};

    \coordinate (R5a) at (0.52,  {0.96*sqrt(3)/2});
    \coordinate (R5b) at (0.512, {0.96*sqrt(3)/2});
    \coordinate (R5c) at (0.524, {0.88*sqrt(3)/2});
    \coordinate (R5d) at (0.56,  {0.7*sqrt(3)/2});
    \coordinate (R5e) at (0.60,  {0.7*sqrt(3)/2});
    \coordinate (R5f) at (0.625, {0.75*sqrt(3)/2});
    \fill[blue!70!black, opacity=0.15] (R5a) -- (R5b) -- (R5c) -- (R5d) -- (R5e) -- (R5f) -- cycle;
    \draw[blue!70!black, thick] (R5a) -- (R5b) -- (R5c) -- (R5d) -- (R5e) -- (R5f) -- cycle;
    \fill[blue!70!black] (R5a) circle (0.5pt);
    \fill[blue!70!black] (R5b) circle (0.5pt);
    \fill[blue!70!black] (R5c) circle (0.5pt);
    \fill[blue!70!black] (R5d) circle (0.5pt);
    \fill[blue!70!black] (R5e) circle (0.5pt);
    \fill[blue!70!black] (R5f) circle (0.5pt);
    \node[blue!70!black, left, font=\small] at (0.50, {0.88*sqrt(3)/2}) {\hs{complexRobot}};
  \end{scope}
\end{tikzpicture}
\caption{Credal sets on the probability simplex.
  \emph{Left:} Ellsberg's urn (\hs{ellsberg}): a line segment at constant $P(\mathsf{R}) = 1 / 3$, with the green/blue split varying freely.
  \emph{Right:} Resulting positions of the \hs{simpleRobot} and \hs{complexRobot} after two moves. While the number of credal vertices is 4 and 16, the hulls are only
  supported by 3 and 6 of these vertices, respectively.}
\label{fig:credal-examples}
\Description{Credal sets shown as shaded regions on probability simplices.}
\end{figure}

Ellsberg's paradox~\cite{ellsberg1961} reveals that people systematically prefer known risks over unknown ones:
given an urn with 30 red balls and 60 that are green or blue in unknown proportion, most people prefer to bet on red.
No single probability distribution can justify this preference~\cite{walley1991}.
The resolution is \emph{imprecise probability}, where uncertainty is modelled not by one distribution but by a \emph{credal set}, a closed,
convex set of distributions~\cite{walley1991,augustin2014}.
Any element of the credal set is considered a plausible probability of the underlying phenomenon, but no single one is privileged.
This captures \emph{Knightian uncertainty}~\cite{walley1991}: irreducible ambiguity that cannot itself be assigned a probability.
In the Ellsberg urn example, the number of red balls is known (so $P(\mathsf{R}) = 1/3$), but the split between green and blue is unknown,
so $P(\mathsf{G})$ and $P(\mathsf{B})$ both range over $[0, 2/3]$.
The associated credal set is shown in Figure~\ref{fig:credal-examples}.
The natural quantities of interest are then \emph{lower} and \emph{upper} probabilities of an event,
obtained by minimizing and maximizing over the credal set.
Convex closure of the set guarantees that these bounds exist, and standard probability machinery (e.g. Bayes' rule and expectations)
generalizes in a canonical way, called the \emph{natural extension}~\cite{walley1991}.
Throughout this paper, we identify a credal set with its extremal vertices since the convex hull recovers the rest.

This paper presents a Haskell DSL for imprecise probabilistic programming.
Programs mix ordinary coin flips with \emph{Knightian} coin flips, whose bias is unknown.
A type-level grading of \emph{Knightian names} tracks the sources of epistemic uncertainty and guarantees \emph{commutativity}: independently specified uncertainties compose without interfering~\cite{liellcock2025}.
Programs compile to binary decision diagrams
(BDDs)~\cite{bryant1986,holtzen2020}, and semiring-parametrized weighted model
counting (WMC) provides exact, differentiable, and interval-bounded inference~\cite{chavira2008}.
In our DSL, the Ellsberg urn is:
\begin{code}
data Ball = Red | Green | Blue

ellsberg :: Imp '["split"] Ball
ellsberg = Imp.do
  isRed   <- flip (1/3)
  isGreen <- knight @"split"
  Imp.return § if isRed then Red else (if isGreen then Green else Blue)
\end{code}
In this program, \hs{flip (1/3)} is an ordinary probabilistic coin flip, while \hs{knight @"split"} is a Knightian coin flip.
The probability of this coin flip lies somewhere in $[0,1]$, but it is not known.
The type-level string \hs{"split"} is a \emph{name} for this source of Knightian uncertainty,
and it appears in the program's type: \hs{Imp '["split"] Ball}.
The type constructor \hs{Imp g a} stands for ``imprecise probabilistic computations of type \hs{a} whose Knightian uncertainty is indexed by the type-level set of names \hs{g}''; we define it formally in~\S\ref{sec:graded-monad}.
The semantics is a graded monad~\cite{katsumata2014,orchard2020}, which tracks the Knightian choices at the type level, following the trace types of Lew et al.~\cite{lew2020}.
GHC's \hs{QualifiedDo} extension provides ergonomic do-notation despite the non-standard bind signature.

Probabilistic programming languages such as Dice~\cite{holtzen2020} already compile coin flips into BDD variables and perform inference via weighted model counting (WMC).
Imprecise probability requires no change to this compilation pipeline.
\textbf{A Knightian coin flip is simply a BDD variable whose weight is not fixed at compilation time but left free}, ranging over $[0,1]$ at inference time.
Exact enumeration resolves the $k$ Knightian variables by visiting their $2^k$ extreme weightings; each extremum is a standard WMC pass over the same compiled BDD, although gradient and interval methods (\S\ref{sec:inference}) avoid this enumeration.

A key subtlety is that the standard monad for sets of distributions (the convex powerset monad) is not commutative.
Its bind permits the resolution of one source of Knightian uncertainty to depend on the outcome of another, so composing two sub-programs in different orders can yield different credal sets.
Following~\cite{liellcock2025}, we use a graded monad, indexed by finite sets of Knightian names, which restores commutativity when sets of names are disjoint.
Our DSL enforces this disjointness at the type level via GHC's type families, so independently specified uncertainties do not interact.
Conversely, dropping names and imposing commutativity unconditionally collapses genuinely different models to the same program (\S\ref{sec:commutativity}).

WMC is a multilinear polynomial in the variable weights~\cite{kimmig2017}, so it is parametric in the semiring, and our implementation exploits this structure.
Dual-number weights~\cite{pearlmutter2017, maene2025} turn a single WMC pass into a gradient computation over the credal set and
interval-valued weights~\cite{Moore1966} yield sound (but not necessarily tight) bounds in one pass.
The result is three inference methods from a single compiled BDD, with no new compilation infrastructure.

In particular, we contribute the following:
\begin{enumerate}
\item A graded monad DSL for imprecise probabilistic programming, where type-level names track Knightian uncertainty and enforce \emph{commutativity} of composition (\S\ref{sec:graded-monad}).
\item The insight that Knightian uncertainty requires no new compilation infrastructure:
  both probabilistic and Knightian choices compile to BDD variables, differing only at inference time (\S\ref{sec:compiling}).
\item A Haskell implementation, inspired by~\cite{kimmig2017}, that obtains three inference methods by instantiating a single weighted-model-counting routine at different semirings:
  exact enumeration, gradient descent over the credal set (\S\ref{sec:ad}), and interval-arithmetic outer approximation (\S\ref{sec:interval-wmc}).
\end{enumerate}

The theoretical foundations of compositional imprecise probability appear in~\cite{liellcock2025}.
This paper is about the Haskell implementation and the interplay between the graded-monad structure and BDD-based inference.
The rest of the paper's layout is as follows. A walkthrough of the language is in \S\ref{sec:examples}.
The implementation details of the DSL are in \S\ref{sec:graded-monad}, and the knowledge compilation to BDDs is in \S\ref{sec:compiling}.
Inbuilt inference methods are in \S\ref{sec:inference}. Two further examples appear in \S\ref{sec:further-examples}, and we discuss related work and conclude in \S\ref{sec:related}.

\section{An Overview of the Language via Examples} \label{sec:examples}

We overview the features of our imprecise probabilistic programming language through an extended example that encodes interval-valued Markov decision processes into our framework,
and demonstrate the alternative inference methods of \S\ref{sec:inference} when enumeration over Knightian variables becomes intractable.
Along the way, we preview combinators such as \hs{interval} and \hs{tag}, with the inference functions that the following sections define in full.
Two further examples, a Monty Hall variant with unknown host bias and an imprecise two-child problem illustrating conditioning, appear in \S\ref{sec:further-examples}.

\subsection{Interval-Valued MDPs} \label{sec:robot-example}

Interval-valued Markov decision processes (IMDPs)~\cite{Givan2000} model sequential decision-making under parameter uncertainty.
Consider a robot that is at one of three points: $P_0$, $P_1$, or $P_2$.
It starts at point $P_0$, and the goal is to get to point $P_2$, which is absorbing.
The probability that the robot successfully moves to the successive point is imprecise and lies in the interval $[0.6, 0.9]$.
\begin{code}
data Position = P0 | P1 | P2

simpleRobot :: Imp '["move1", "move2"] Position
simpleRobot = Imp.do
  move1 <- interval @"move1" 0.6 0.9
  let pos1 = step P0 move1
  move2 <- interval @"move2" 0.6 0.9
  Imp.return § step pos1 move2
\end{code}
The function \hs{step} advances the robot based on its move, and the command \hs{interval} is a derived operator and uses one Knightian name to produce an imprecise coin flip between the provided bounds.
The two Knightian sources dictate that there are $2^2 = 4$ valuations to compute to find the credal set of the robot's final state.
The distribution for each Knightian valuation, represented as pairs of a value and probability, can be produced using the following method.
\begin{code}
credalVertices :: Ord a => Imp g a -> [[(a, Double)]]
\end{code}
In our example, running \hs{credalVertices simpleRobot} yields the following output.

\begin{center}
\begin{tabular}{cc|ccc}
\toprule
\hs{move1} & \hs{move2} & $P(P_0)$ & $P(P_1)$ & $P(P_2)$ \\
\midrule
0.6 & 0.6 & 0.16 & 0.48 & 0.36 \\
0.6 & 0.9 & 0.04 & 0.42 & 0.54 \\
0.9 & 0.6 & 0.04 & 0.42 & 0.54 \\
0.9 & 0.9 & 0.01 & 0.18 & 0.81 \\
\bottomrule
\end{tabular}
\end{center}

Each row of the table corresponds to one valuation of the two Knightian variables at their extreme biases.
The first two columns record the extreme probabilities of each successful move,
and the remaining columns give the resulting probability of the robot ending in each position.
The intervals are symmetric, so the valuations $\{\hs{move1} \mapsto 0.6, \hs{move2} \mapsto 0.9\}$ and $\{\hs{move1} \mapsto 0.9, \hs{move2} \mapsto 0.6\}$
produce the same distribution, leaving three distinct vertices.
The credal set, shown in Figure~\ref{fig:credal-examples}, is a triangle in the simplex over $P_0$, $P_1$ and $P_2$, the convex hull of the three independent points.

\subsubsection{Robot Navigation with Unknown Dynamics}\label{sec:unknown-dynamics}

To illustrate the framework's compositionality, we extend the example.
The robot may also move backwards, and the imprecise dynamics that govern whether the robot
moves forwards, remains stationary, or moves backwards are given by an external function \hs{robotDynamics}.
\begin{code}
data Move = Backwards | Stationary | Forwards

complexRobot = Imp.do
  move1 <- tag @"move1" robotDynamics
  let pos1 = step P0 move1
  move2 <- tag @"move2" robotDynamics
  Imp.return § step pos1 move2
\end{code}
We don't specify the type of \hs{complexRobot} because it is dependent on the type of \hs{robotDynamics}.
This is not an issue, because Haskell's type system is strong enough to infer its type once \hs{robotDynamics} is specified.
Additionally, each of the calls to the robot's dynamics is tagged with a symbol, because the names of the Knightian choices dictate whether they are correlated.
The \hs{tag} function regrades the names from the robot's dynamics for each move, so they are distinct.
One possible implementation of the robot dynamics is given below, and the resulting
credal set of the robot's final position is shown in Figure~\ref{fig:credal-examples}.
\begin{code}
robotDynamics :: Imp '["b", "f"] Move
robotDynamics = Imp.do
  goForwards  <- interval @"f" 0.5 0.8
  goBackwards <- interval @"b" 0.0 0.2
  Imp.return § if goForwards then Forwards
               else if goBackwards then Backwards
               else Stationary 
\end{code}
The number of vertices required for inference has increased drastically. Each move introduces two Knightian choices, meaning
there are $2^{2\times 2} = 16$ valuations. While $16$ valuations are still straightforward to compute,
we have additional inference methods that do not require enumerating over all of them, one of which we preview now.

\subsubsection{Gradient Descent on the Credal Set}

Now consider that the example was an abstraction of a robot crossing a road.
That is, $P_0$ is the starting side of the road, $P_1$ is the middle of the road, and $P_2$ is the other side of the road (the goal).
The robot should not only reach the goal, but also avoid ending in the middle of the road, where it could be hit by oncoming traffic.
We assign a score to each final position, rewarding the robot for reaching the other side of the road, and penalizing it for stopping in the middle.
\begin{code}
score :: Position -> Double
score P0 = 0
score P1 = -10
score P2 = 10
\end{code}
Using \emph{dual numbers} for the variable weights in WMC permits the same algorithm to compute forward-mode derivatives over the Knightian variables in the BDD.
Therefore, we can perform gradient ascent (or descent) over the credal set to calculate the best (or worst) expected score of our imprecise probabilistic program.

The interface to the credal optimization takes in a maximum number of learning steps (it will early-abort if converged),
a learning rate, a scoring function, and an \hs{Imp} program, respectively.
\begin{code}
credalOptimizeExpectation :: Ord a => Int -> Double -> (a -> Double)
  -> Imp g a -> ([(String, Double)], Double)
\end{code}
The output is a list of name and probability pairs, encoding the probability of each Knightian choice that optimizes the scoring function, along with the resulting score.
In the robot example, we may use this function to search for the worst possible expected score given the four Knightian inputs, \hs{credalOptimizeExpectation 200 (-0.1) score complexRobot},
producing:

\begin{center}
\begin{tabular}{cccc|c}
\toprule
\hs{"move1.b"} & \hs{"move1.f"} & \hs{"move2.b"} & \hs{"move2.f"} & Score \\
\midrule
$0.0$ & $1.0$ & $0.0$ & $0.0$ & $4.5$ \\
\bottomrule
\end{tabular}
\end{center}

At each of the robot's moves, increasing the Knightian distribution labelled by \hs{"f"} increases the chance of the robot moving forwards,
while increasing the Knightian distribution labelled by \hs{"b"} increases the chance of the robot moving backwards.
We can see from the results that the worst score happens when the robot is most likely to move forwards on the first move, and then
most likely to move in neither direction on the second move. However, this still results in a positive minimum expected score of $4.5$.

\section{The Graded Monad DSL} \label{sec:graded-monad}

The central data type underlying the DSL is \hs{Imp}, a \emph{generalized algebraic data type} (GADT).
GADTs extend ordinary algebraic data types by letting each constructor refine the parameters in its return type,
permitting the constructors below to specify different grades.
\hs{Imp} is parametrized by an output type~\hs{a}, and a type-level set of symbols~\hs{g} that represents the grade.
\begin{code}
data Imp (g :: [Symbol]) a where
  ImpReturn  :: a -> Imp '[] a
  ImpBind    :: Imp g1 a -> (a -> Imp g2 b) -> Imp (Merge g1 g2) b
  ImpFlip    :: !Double -> Imp '[] Bool
  ImpKnight  :: KnownSymbol n => Proxy n -> Imp '[n] Bool
  ImpObserve :: Bool -> Imp '[] ()
  ImpTag     :: KnownSymbol t => Proxy t -> Imp g a -> Imp (TagAll t g) a
\end{code}
We represent finite sets of symbols canonically as strictly ordered type-level lists.
Grades are introduced only as empty or singletons, and the type families \hs{Merge} and \hs{TagAll}
(defined in \S\ref{sec:type-machinery}) preserve the ordering invariant.
In informal terms, each constructor has the following meaning.
\begin{itemize}
\item \hs{ImpReturn} injects a value into a pure computation. Its grade is empty~\hs{'[]} because a pure computation has no Knightian uncertainty.

\item \hs{ImpBind} sequences two computations. The first has grade \hs{g1} and the result is passed to
  the continuation, which produces a computation at grade~\hs{g2}. The resulting computation has grade~\hs{Merge g1 g2}, where \hs{Merge} is
  a type-level function that computes the disjoint union of two sorted name lists (raising an error on overlap).

\item \hs{ImpFlip} is a probabilistic coin flip with a known bias. It takes a value in $[0,1]$ that parametrizes the
  bias of the coin flip, and returns a \hs{Bool} at grade~\hs{'[]}. A coin flip with known bias has no Knightian uncertainty.

\item \hs{ImpKnight} is the Knightian counterpart to \hs{ImpFlip}.
  It represents a binary choice for which we do not know the distribution.
  It introduces a name~\hs{n} that tracks the Knightian input.
  The name is represented by the type \hs{Proxy n} whose unique value \hs{Proxy} carries no run-time data,
  but permits passing the type-level name~\hs{n} as an argument.
  The constraint \hs{KnownSymbol n} says that~\hs{n} can be reflected back to a \hs{String} at run-time to obtain the Knightian variable's name.
  Coin flips that share the same name share the same unknown distribution.
  The resulting computation only depends on a single Knightian choice and thus has singleton grade~\hs{'[n]}.

\item \hs{ImpObserve} conditions on a boolean (see \S\ref{sec:twochild}), restricting to executions where the predicate holds.

\item \hs{ImpTag} renames each of the Knightian names by attaching a tag, using a second type family \hs{TagAll t g} that prepends the symbol~\hs{t}
  to every name in~\hs{g}.
  This functionality is crucial for compositionality. It is required for multiple calls to the same external imprecise probabilistic program,
  as seen in the \hs{complexRobot} example in \S\ref{sec:unknown-dynamics}.
\end{itemize}

\paragraph{Binary primitives.}
The core DSL exposes only binary primitives for probabilistic and Knightian choices.
Any finite distribution can be encoded as a sequence of biased coin flips, the standard way to implement categoricals in discrete probabilistic languages such as Dice~\cite{holtzen2020}.
A bit-level encoding also works for Knightian choices over finite values, with the credal set recovered as the convex closure of the valuations.
We could expose non-binary \hs{flip} and \hs{knight} primitives directly,
but the mechanics of inference remain the same since any such choice reduces to binary at the BDD compilation step (\S\ref{sec:compiling}).
With dynamic variable ordering during BDD compilation, the overhead of this encoding is largely absorbed~\cite{kimmig2017,holtzen2020}.

\subsection{Smart Constructors and Intervals} \label{sec:smart-const}

The \hs{ImpKnight} and \hs{ImpTag} constructors require a \hs{Proxy}.
To overcome this inconvenience, we provide smart constructors \hs{knight} and \hs{tag} that use a visible type application instead:
\begin{code}
knight :: forall n. KnownSymbol n => Imp '[n] Bool
knight = ImpKnight (Proxy @n)

tag :: forall t g a. KnownSymbol t => Imp g a -> Imp (TagAll t g) a
tag = ImpTag (Proxy @t)
\end{code}
These enable the \hs{knight @"split"} and \hs{tag @"move1"} patterns from the earlier examples.
For the remaining constructors, such as \hs{ImpFlip} and \hs{ImpObserve}, we provide thin wrappers \hs{flip} and \hs{observe}.

Another useful derived combinator is \hs{interval}, which models a coin whose bias is bounded:
\begin{code}
interval :: forall n. KnownSymbol n => Double -> Double -> Imp '[n] Bool
interval lo hi = ImpBind (ImpKnight (Proxy @n)) § λb ->
  ImpFlip (if b then hi else lo)
\end{code}
The idea is that \hs{interval @"n" lo hi} returns \hs{True} with some probability in the range~$[\hs{lo}, \hs{hi}]$.
Its implementation makes a Knightian choice between the coin flip biases of~\hs{lo} and~\hs{hi}, which semantically is the closed interval between them.
Then a probabilistic flip with the chosen bias resolves the outcome.
The \hs{simpleRobot} example in \S\ref{sec:robot-example} uses \hs{interval @"move1" 0.6 0.9} to model the robot's imprecise dynamics.

\subsection{Names and Correlation} \label{sec:correlation}

At the heart of the DSL is the role played by the names of the Knightian choices.
Knightian choices with the \emph{same} name are \emph{correlated}.
They must be resolved using the same bias.
Choices with \emph{different} names are \emph{independent}, and there is no relation in the way they are resolved.

\noindent
\begin{minipage}[t]{0.48\linewidth}
\begin{code}
data Three = R | G | B

dependent :: Imp '["a1"] Three
dependent = Imp.do
  x <- flip 0.5
  if x then Imp.do
    y <- knight @"a1"
    Imp.return (if y then R else G)
  else Imp.do
    y <- knight @"a1"
    Imp.return (if y then R else B)
\end{code}
\end{minipage}
\hfill
\begin{minipage}[t]{0.48\linewidth}
\begin{code}


independent :: Imp '["a1", "a2"] Three
independent = Imp.do
  x <- flip 0.5
  if x then Imp.do
    y <- knight @"a1"
    Imp.return (if y then R else G)
  else Imp.do
    y <- knight @"a2"
    Imp.return (if y then R else B)
\end{code}
\end{minipage}
\medskip

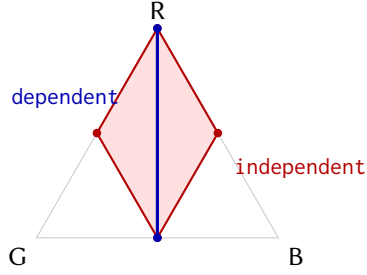
\begin{figure}[t]
\centering
\begin{tikzpicture}[scale=3.2]
  \coordinate (G) at (0, 0);
  \coordinate (B) at (1, 0);
  \coordinate (R) at (0.5, {sqrt(3)/2});
  \draw[gray!40] (G) -- (B) -- (R) -- cycle;
  \node[below left]  at (G) {$\mathsf{G}$};
  \node[below right] at (B) {$\mathsf{B}$};
  \node[above]       at (R) {$\mathsf{R}$};
  \coordinate (TT) at (0.5, {sqrt(3)/2});                
  \coordinate (TF) at ({0.5+0.5*0.5}, {0.5*sqrt(3)/2}); 
  \coordinate (FF) at (0.5, 0);                           
  \coordinate (FT) at ({0+0.5*0.5}, {0.5*sqrt(3)/2});    
  \fill[red!20, opacity=0.6] (TT) -- (TF) -- (FF) -- (FT) -- cycle;
  \draw[red!70!black, thick] (TT) -- (TF) -- (FF) -- (FT) -- cycle;
  \fill[red!70!black] (TT) circle (0.5pt);
  \fill[red!70!black] (TF) circle (0.5pt);
  \fill[red!70!black] (FF) circle (0.5pt);
  \fill[red!70!black] (FT) circle (0.5pt);
  \node[red!70!black, right, font=\small] at (0.78, {sqrt(3)/6}) {\hs{independent}};
  \coordinate (f1b0) at (0.5, {sqrt(3)/2});          
  \coordinate (f1b1) at (0.5, 0);                     
  \draw[blue!70!black, very thick] (f1b0) -- (f1b1);
  \fill[blue!70!black] (f1b0) circle (0.5pt);
  \fill[blue!70!black] (f1b1) circle (0.5pt);
  \node[blue!70!black, left, font=\small] at (0.38, {2*sqrt(3)/6}) {\hs{dependent}};
\end{tikzpicture}
\caption{Credal sets on the probability simplex over $\{\mathsf{R},\mathsf{G},\mathsf{B}\}$.
  The semantics of \hs{dependent} (one Knightian name) is a line segment.
  The semantics of \hs{independent} (two Knightian names) is a quadrilateral.
  Increasing the number of independent Knightian choices enlarges the credal set.}
\label{fig:credal-names}
\Description{Credal sets for the named-choice programs on the three-outcome simplex.}
\end{figure}

To illustrate, consider the above two programs adapted from~\cite{liellcock2025}, both returning a value in
$\{\mathsf{R}, \mathsf{G}, \mathsf{B}\}$.
In the first, a Knightian choice is made in each code branch, but they share the same underlying distribution labelled~\hs{"a1"}.
If the probability of $\true$ is~$1$, then the result of the computation will be~\hs{R}.
On the other extreme, if the probability of $\true$ is~$0$, then the result of the computation will be equal odds between~\hs{G} and~\hs{B}.
The hidden distribution~\hs{"a1"} can be any convex combination of these two extremes, so the resulting credal set is the line segment between these
two points in the probability simplex, as shown in Figure~\ref{fig:credal-names}.
The result cannot be equal odds between~\hs{R} and~\hs{G}, because the two Knightian choices are correlated.

The program \hs{independent} is almost identical.
The only structural difference is that the second branch uses a different underlying distribution with name~\hs{"a2"} for its Knightian choice.
But the semantics differ because the Knightian choices in each branch are now resolved independently. The resolution of~\hs{"a1"} does not constrain~\hs{"a2"}, or vice versa.
It is now possible for the result to be equal odds between~\hs{R} and~\hs{G},
because distribution~\hs{"a1"} can resolve to certainty of $\false$ while distribution~\hs{"a2"} can resolve to certainty of $\true$.
There are four possible resolutions of the imprecise probability in total:
\[
\left\{\mathsf{R},  \
\frac{1}{2}\mathsf{R} + \frac{1}{2}\mathsf{G}, \
\frac{1}{2}\mathsf{R} + \frac{1}{2}\mathsf{B}, \
\frac{1}{2}\mathsf{G} + \frac{1}{2}\mathsf{B}\right\}
\]
The credal set is the convex hull of these four vertices, a diamond strictly larger than the line segment, as shown in Figure~\ref{fig:credal-names}.

This distinction is clear from the type system.
The grade of \hs{dependent} is~\hs{'["a1"]}, recording one degree of Knightian uncertainty.
On the other hand, the grade of \hs{independent} is~\hs{'["a1", "a2"]}, recording two degrees of Knightian uncertainty.

When programs are composed via bind, the name sets are combined by disjoint union,
reflecting the treatment of each Knightian name as a distinct source of epistemic uncertainty.
If two sub-programs reused the same name, the resulting program would silently correlate them.
Duplicate names on the same code path, as in the following code, are ambiguous between
\emph{reuse the same sample} and \emph{resample with the same underlying bias}.
\begin{code}
  y <- knight @"a1"
  z <- knight @"a1"
\end{code}
We eliminate this ambiguity by enforcing disjointness and requiring explicit renaming (e.g. with \hs{tag}) when independence is intended.

\subsection{Commutativity} \label{sec:commutativity}

The examples above show that sub-programs with disjoint name sets behave independently.
The underlying property is \emph{commutativity}~\cite{liellcock2025}:
when the set of names of two sub-programs is disjoint, the order of composition does not affect the credal set.

\noindent
\begin{minipage}[c]{0.48\linewidth}
\begin{code}
Imp.do
  x <- p
  y <- q
  f x y
\end{code}
\end{minipage}
\hfill$=$\hfill
\begin{minipage}[c]{0.48\linewidth}
\begin{code}
Imp.do
  y <- q
  x <- p
  f x y
\end{code}
\end{minipage}

\noindent
where \hs{p :: Imp g1 a} and \hs{q :: Imp g2 b} have disjoint grades.

The standard monad for sets of distributions (the convex powerset monad) is not commutative.
Composing unknowns in different orders can yield different credal sets, because the monad permits arbitrary correlations between them.
Our graded monad avoids this by using names to separate independent sources of uncertainty.
Conversely, if we dropped names and required commutativity unconditionally,
\hs{dependent} and \hs{independent} (\S\ref{sec:correlation}) would collapse to the same program.
The full derivation appears in~\cite[Fig.~2]{liellcock2025}:
commutativity and if-then-else hoisting rewrite \hs{dependent} so that each branch of the conditional draws its own \hs{knight}.

A recent extension to Dice, \emph{noDice}~\cite{gurtler2026}, takes an alternative route.
It adds an unnamed non-deterministic flip command \hs{nflip} and provides state-of-the-art inference algorithms.
However, the lack of names constrains the command to be local, so common program equivalences, such as commutativity above, no longer hold.
Our names, by contrast, reinstate these program equivalences and permit a single unified compilation scheme for a range of inference methods (\S\ref{sec:inference}).

\subsection{Library Combinators} \label{sec:library-combinators}

The core constructors of \hs{Imp} are sufficient for expressing any imprecise probabilistic program of interest in this paper.
In practice, they are tedious to use for writing large programs with many uncertainties.
We therefore expose a layer of derived combinators built on top of this core to promote easy encoding of imprecise systems into our language.
Each one elaborates to the \hs{Imp} constructors at compile time with no run-time cost or semantic primitives.
They let programs with many uncertainties be written concisely.
We give a functional overview of the main combinators and omit their implementation.
\begin{itemize}
\item \hs{intervalMap} and \hs{knightMap} populate a type-level list of names with independent interval or Knightian choices.
  \hs{intervalN} and \hs{knightN} generate a fixed number of choices based on a type-level integer input.
\begin{code}
intervalMap @'["a1", "a2"] 0.5 0.9 :: Imp '["a1", "a2"] [Bool]
\end{code}

\item \hs{tagMap} and \hs{tagN} are numbered and iterable versions of \hs{tag}.
\begin{code}
tagN @2 @"move" robotDynamics
  :: Imp '["move1.b", "move1.f", "move2.b", "move2.f"] [Move]
\end{code}

\item \hs{foldN}, \hs{scanN}, and \hs{foldmN} are numbered recursive schemes, accumulating the result at each step.
  The monadic fold, \hs{foldmN}, can only be over programs that do not introduce new Knightian uncertainty.
  They permit expressing the \hs{complexRobot} example with a single line.
\begin{code}
complexRobot = foldN @2 @"move" robotDynamics P0 step
\end{code}
\end{itemize}

\subsection{Type-Level Machinery} \label{sec:type-machinery}

The grading monoid operations are implemented as type families:
\begin{code}
type family Merge (xs :: [Symbol]) (ys :: [Symbol]) :: [Symbol] where
  Merge '[]       ys        = ys
  Merge xs        '[]       = xs
  Merge (x ': xs) (y ': ys) = MergeH (CmpSymbol x y) x xs y ys

type family MergeH (o :: Ordering) (x :: Symbol) (xs :: [Symbol])
                   (y :: Symbol) (ys :: [Symbol]) :: [Symbol] where
  MergeH 'LT x xs y ys = x ': Merge xs (y ': ys)
  MergeH 'EQ x _  _ _  =
    TypeError ('Text "Duplicate Knightian name: " ':<>: 'ShowType x)
  MergeH 'GT x xs y ys = y ': Merge (x ': xs) ys
\end{code}
\hs{Merge} is a type-level, order-preserving merge of two ordered lists, using \hs{MergeH} as a helper.
The disjointness check is performed here. We enforce the invariant that the grading list is sorted by only introducing empty lists
or singletons in the core DSL, and only combining them with \hs{Merge}. So \hs{Merge} will always eventually compare and catch duplicate entries.
If a programmer writes a program that binds two sub-programs both using the name~\hs{"a"}, GHC emits a helpful type error:

\begin{plain}
  Duplicate Knightian name: "a"
\end{plain}
This design enforces that grade-disjointness checking happens at compile time.
The grades are erased at run-time, leaving \hs{Imp} a free monad-like syntax tree waiting to be interpreted.
The names carry no run-time cost, but they still provide a static guarantee about the compositional structure.

\subsection{Graded Monad Laws and Partiality}

\emph{Graded monads}~\cite{katsumata2014} are a generalization of monads that permits annotation of computational effects from a monoid or monoidal category.
Return produces a computation at the unit annotation and bind tensors annotations together.

The type-level ordered lists of \hs{Imp} encode the monoid of finite sets of symbols, $(P_{f}(\hs{Symbol}), \cup, \emptyset)$.
Semantically, \hs{Imp g a} is the graded monad $T_g(a) = (2^g \to D(a))$~\cite{liellcock2025};
a function from valuations of the Knightian names $g$ to distributions over $a$.
Return is the trivial valuation, bind partitions the valuations across the two sub-computations, and
tagging is a special type of regrading along the bijective function that prepends a symbol to each name.

At the implementation level, the type family \hs{Merge} agrees with union on disjoint sets and throws a type error otherwise.
This restriction deliberately prevents silent correlations between programs that overlap on Knightian names.
So \hs{Imp} satisfies the graded monad laws modulo the disjointness side condition.

An alternative design avoiding partiality would be to use a tagged disjoint union for \hs{Merge}.
However, identical names on different code paths may receive different tags,
silently decoupling them and breaking the deliberate dependence mechanism in \S\ref{sec:correlation}.

\subsection{Qualified Do-Notation}

The graded monad operations of \hs{Imp} have the following types:

\begin{code}
return :: a -> Imp '[] a
return = ImpReturn

(>>=) :: Imp g1 a -> (a -> Imp g2 b) -> Imp (Merge g1 g2) b
(>>=) = ImpBind
\end{code}
These are not the standard types of the Haskell monad class.
The grades (and therefore the functors) of the two input computations to~\verb|(>>=)| differ.
The \hs{return} operation is also constrained to produce a value under the functor~\hs{Imp '[]}.
GHC's \hs{QualifiedDo} extension resolves this mismatch by permitting us to write \hs{Imp.do} instead of \hs{do}.
This instructs GHC to desugar \hs{<-} using the qualified name \verb|Imp.>>=|, rather than the \hs{Prelude} version.
The result is that programs in our DSL look like ordinary Haskell, as seen in the previous examples.

\paragraph{GHC extensions.}
\begin{sloppypar}
The library is written against \hs{GHC2021}, which has many of the required extensions like \hs{TypeApplications} and \hs{TypeOperators} built in.
Beyond this, the DSL requires \hs{DataKinds} and \hs{TypeFamilies} for the type-level
name lists and \hs{Merge} and \hs{TagAll} families, \hs{GADTs} for the \hs{Imp} type,
and \hs{QualifiedDo} and \hs{RebindableSyntax} for ergonomic do-notation.
Internally, the library also uses \hs{UndecidableInstances} (for the recursive
type families) and \hs{AllowAmbiguousTypes} (for the smart constructors of \S\ref{sec:smart-const}).
User code requires only \hs{DataKinds}, \hs{QualifiedDo} and \hs{RebindableSyntax}.
\end{sloppypar}

\section{Compiling to BDDs}\label{sec:compiling}

While the~\hs{Imp} DSL is convenient for producing readable imprecise probabilistic programs,
it is an inefficient representation for inference. We use \emph{knowledge compilation}~\cite{Darwiche2002KC, bowers2025} to reduce it
to a computable form.
The simple and compositional compilation strategy is identical to that of a precise probabilistic language.
An \hs{Imp g a} program is compiled to a list of \emph{worlds},
\begin{code}
type World a = (a, BDD)
\end{code}
that consists of a return value~\hs{a} and a binary decision diagram \hs{BDD}.
A BDD is an efficient encoding of a boolean formula,
which captures the valuations of the probabilistic coin flips and Knightian choices that return the given value.
The results of compiling programs \hs{dependent} and \hs{independent} for value \hs{R} are shown in Figure~\ref{fig:bdd}.

The top-level compiler has the signature,
\begin{code}
compile :: Ord a => Imp g a -> ([World a], CompileState)
\end{code}
that returns a list of worlds and a \hs{CompileState}. This holds the BDD manager (the hash-consing table for BDD nodes),
the weight assignment for probabilistic variables, and the identities of Knightian variables:
\begin{code}
data CompileState = CompileState
  { csMgr        :: !BDDManager
  , csWeights    :: !(WMCParams RealS)
  , csKnightVars :: ![(String, VarLabel)]
  , csProbVars   :: ![VarLabel]
  }
\end{code}

\subsection{The BDD Representation}

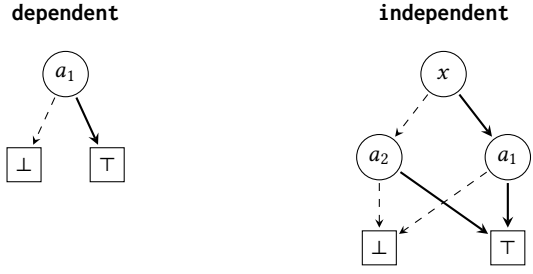
\begin{figure}[t]
\centering
\begin{tikzpicture}[
  bvar/.style={circle, draw, minimum size=6mm, inner sep=0pt, font=\small},
  bterm/.style={draw, minimum size=4.5mm, inner sep=1pt, font=\small},
  high/.style={thick,        ->, >=stealth, shorten >=1pt, shorten <=1pt},
  low/.style ={dashed,       ->, >=stealth, shorten >=1pt, shorten <=1pt},
]
  \begin{scope}[shift={(-2.5, 0)}]
    \node[font=\small\bfseries] at (0, 0.8) {\hs{dependent}};
    \node[bvar]  (a1)   at (0, 0)        {$a_1$};
    \node[bterm] (zero) at (-0.55, -1.2) {$\bot$};
    \node[bterm] (one)  at ( 0.55, -1.2) {$\top$};
    \draw[low]  (a1) -- (zero);
    \draw[high] (a1) -- (one);
  \end{scope}

  \begin{scope}[shift={(2.5, 0)}]
    \node[font=\small\bfseries] at (0, 0.8) {\hs{independent}};
    \node[bvar]  (x)    at (0, 0)        {$x$};
    \node[bvar]  (a2)   at (-0.85, -1.1) {$a_2$};
    \node[bvar]  (a1)   at ( 0.85, -1.1) {$a_1$};
    \node[bterm] (zero) at (-0.85, -2.3) {$\bot$};
    \node[bterm] (one)  at ( 0.85, -2.3) {$\top$};
    \draw[low]  (x)  -- (a2);
    \draw[high] (x)  -- (a1);
    \draw[low]  (a2) -- (zero);    
    \draw[high] (a2) -- (one);     
    \draw[low]  (a1) -- (zero);    
    \draw[high] (a1) -- (one);     
  \end{scope}
\end{tikzpicture}
\caption{BDDs for the value~\hs{R} compiled from \hs{dependent} and \hs{independent}.
  Solid arrows go to the variable's \emph{high} child and dashed arrows to the \emph{low} child.
  In \hs{dependent}, the two paths producing \hs{R} share the Knightian variable~$a_1$, collapsing the BDD to a single variable.
  In \hs{independent}, the two paths use distinct Knightian variables and thus remain separate.}
\label{fig:bdd}
  \Description{Two binary decision diagrams.}
\end{figure}

We use a standard BDD representation with hash-consing and complemented edges~\cite{bryant1986,brace1990}.
A BDD is either a constant boolean or a reference to a node in the unique table or its complement:
\begin{code}
data BDD
  = BDDTrue
  | BDDFalse
  | BDDRef  !NodeId
  | BDDComp !NodeId
\end{code}
Each node records a variable label and two children:
\begin{code}
data BDDNode = BDDNode
  { bddVar  :: !VarLabel
  , bddLow  :: !BDD
  , bddHigh :: !BDD
  }
\end{code}
Complemented edges give us $\mathcal{O}(1)$ negation because the function merely flips between \hs{BDDRef} and \hs{BDDComp},
and halves the size of the lookup table, since we normalize so that the low edge is never complemented.
All BDD operations (conjunction, disjunction, if-then-else) are implemented via the standard recursive ITE algorithm with a computed-table cache~\cite{brace1990, bryant1986}.

Our BDDs are reduced and ordered. There is a fixed total order on the variables, given by creation order during compilation,
used by the ITE operator to decide which variable to branch on next.
Variable ordering can affect performance~\cite{Rudell1993, Panda1995}, and even correctness when non-determinism and probabilities are combined,
because maxing and summing do not commute~\cite{cho2025}.
Our implementation avoids the variable-order restriction by fixing the Knightian valuations \emph{externally}, reducing each WMC pass to a summative computation.
So the choice of variable ordering only affects BDD size, not correctness, and we are free to employ reordering heuristics, such as sifting.
As these are primarily performance optimizations, they are beyond the scope of this pearl.

\paragraph{Compilation rules.}
The compiler is a straightforward structural recursion on the \hs{Imp} syntax tree, running in a state monad that owns the \hs{CompileState}.
It employs two helper functions for allocating BDD variables.
\hs{flipVar p} allocates a fresh probabilistic BDD variable, records its weight $(1{-}p, p)$ in \hs{csWeights}, and returns the variable's BDD.
\hs{knightVar @n} returns the BDD for the Knightian name~\hs{n}, allocating a fresh variable the first time the name is seen and reusing the existing one thereafter.
We provide the main rules to give intuition for how compilation proceeds.

\begin{description}
\item[\hs{ImpReturn a}:] A pure value is always available, guarded by $\top$:
\begin{code}
compileM (ImpReturn a) =
  return [(a, BDDTrue)]
\end{code}

\item[\hs{ImpFlip p}:] A probabilistic coin flip creates a fresh weighted BDD variable~$v$ and returns two worlds,
  one where the coin lands heads (guarded by $v$) and one where it lands tails (guarded by $\lnot v$).
\begin{code}
compileM (ImpFlip p) = do
  v <- flipVar p
  return [(True, v), (False, bddNot v)]
\end{code}

\item[\hs{ImpKnight (\_ :: Proxy n)}:] A Knightian choice compiles similarly, except \hs{knightVar} consults \hs{csKnightVars}.
  If the name~\hs{n} has been seen before, the existing variable is reused. Otherwise a fresh, unweighted BDD variable is allocated for \hs{n}.
\begin{code}
compileM (ImpKnight (_ :: Proxy n)) = do
  v <- knightVar @n
  return [(True, v), (False, bddNot v)]
\end{code}
This is the key insight of the compilation: both \hs{flip} and \hs{knight} produce the same kind of object, a BDD variable that splits the outcomes in two, but they differ in how they are interpreted.
Variables from probabilistic flips get weights and contribute to the weighted model count,
while Knightian variables are left unweighted. Their interpretation depends on what type of inference is performed later.
The BDD is agnostic to the kind of uncertainty the variables represent.
The same-name reuse in \hs{knightVar} is also what causes \hs{dependent} and \hs{independent} to compile to structurally different BDDs (Figure~\ref{fig:bdd}).

\item[\hs{ImpBind m f}:] Monadic bind compiles the computation~\hs{m} to its guarded worlds,
then for each world $(a, g)$, it compiles the continuation \hs{f a} and conjoins the guards, $g \wedge g'$,
expressing ``$m$ returned $a$ \emph{and} the continuation returned $b$''.
\begin{code}
compileM (ImpBind m f) = do
  mWorlds <- compileM m
  fmap concat § mapM (λ(a, g) -> do
    fWorlds <- compileM (f a)
    mapM (λ(b, g') -> do
      gAnd <- bddAnd g g'
      return (b, gAnd)) fWorlds) mWorlds
\end{code}

\item[\hs{ImpObserve b}:] Conditioning produces a single world guarded by the given boolean.
\begin{code}
compileM (ImpObserve b) =
  return [((), if b then BDDTrue else BDDFalse)]
\end{code}
Subsequent computations are conjoined with this guard via bind.
This conditions on the boolean value because any world whose guard is unsatisfiable under the observation will have probability zero.
Renormalization happens at inference, as discussed further in \S\ref{sec:wmc}.

\end{description}

\subsection{Weighted Model Counting} \label{sec:wmc}

Given a compiled BDD and a weight assignment, we compute probabilities via \emph{weighted model counting} (WMC)~\cite{darwiche2002,chavira2008}.
The WMC of a BDD is defined by recursion,
\[
  \WMC(\top) = 1, \qquad
  \WMC(\bot) = 0, \qquad
  \WMC(v \mapsto (l, h)) = w^\bot_v \cdot \WMC(l) + w^\top_v \cdot \WMC(h)\text,
\]
where $v\mapsto (l,h)$ represents an internal node with variable $v$, low child $l$ and high child $h$,
and $w^\bot_v$ and $w^\top_v$ are the weights for $v$ being false and true, respectively.
For a flip variable with parameter~$p$, these are $1{-}p$ and $p$.

We parametrize this computation over an abstract \emph{semiring}:
\begin{code}
class Semiring a where
  zero, one :: a
  (.+.), (.*.) :: a -> a -> a
\end{code}
The WMC function is then polymorphic in the semiring:
\begin{code}
wmc :: Semiring s => BDDManager -> WMCParams s -> BDD -> s
\end{code}
The implementation walks the BDD top-down with memoization, handling complemented edges by negating children.
It is a textbook BDD-based WMC algorithm, but the semiring abstraction allows WMC over different parameters,
which comes into play when implementing automatic differentiation in \S\ref{sec:ad} and efficient conservative inference in \S\ref{sec:interval-wmc}.

\paragraph{Renormalization.} Operationally, \hs{observe} introduces an \emph{evidence guard} that assigns a weight of zero to any inconsistent world.
During inference, valuations of Knightian choices are processed independently. If the WMC of the valuation is zero, it is infeasible and discarded.
Otherwise, posterior quantities are normalized on a per-valuation basis. There is no \emph{global} renormalization, only local probabilistic scaling.
This implements the \emph{natural extension}~\cite{walley1991} of conditioning to imprecise probabilities.
We work through a concrete instance of an imprecise conditioning problem in \S\ref{sec:twochild}.

\section{Inference on BDDs} \label{sec:inference}

For a purely probabilistic program with no Knightian uncertainty, computing marginals is a single WMC call per world.
\begin{code}
preciseMarginal :: Ord a => Imp '[] a -> [(a, Double)]
\end{code}
The graded monad gives a layer of safety here.
The user cannot accidentally input a program containing Knightian uncertainty to calculate its \hs{preciseMarginal}, because the typechecker will reject it.

For programs with Knightian variables~\hs{Imp g a}, we compute interval-valued probabilities by enumerating
all $2^{|g|}$ boolean valuations of the Knightian variables and running WMC for each, taking the
minimum and maximum:
\begin{code}
intervalProbability :: (Ord a) => Imp g a -> (a -> Bool) -> (Double, Double)
\end{code}
Each valuation is handled by conditioning the BDD, restricting the Knightian variables to a bias of 0 or 1, and then ordinary WMC is applied to the result.
The enumeration is exponential in the number of Knightian variables, but there is no way to get around this exponential growth,
because computing tight bounds on imprecise probabilities is \#P-hard in general~\cite{maua2018}.
In practice, many models have few Knightian variables, each representing a dimension of epistemic uncertainty, so enumeration is feasible.
However, there are cases where this is not sufficient,
so we provide alternative approaches for inference when the number of Knightian variables is significant.
A comparison of our inference tools is given in \S\ref{sec:performance}.

\subsection{Automatic Differentiation for Credal Optimization} \label{sec:ad}

The semiring abstraction introduced in \S\ref{sec:wmc}, following Kimmig et al.~\cite{kimmig2017}, also yields exact gradients.
Rather than using the semiring over the real numbers, we can use \emph{dual numbers}, which pair a value with a vector of partial derivatives.
\begin{code}
data DualS = DualS !Double !(V.Vector Double)
\end{code}
Dual numbers form a semiring, allowing reuse of the WMC algorithm to obtain the exact gradients of probabilities with respect to model parameters.
The semiring instance implements the standard rules of forward-mode automatic differentiation (AD).
Addition propagates partials by summation, while multiplication uses the product rule, $\frac{d}{d\theta}(a \cdot b) =
a \cdot \frac{db}{d\theta} + \frac{da}{d\theta} \cdot b$.
Forward-mode AD is then automatically done through the inference pipeline.
\begin{code}
instance Semiring DualS where
  zero = DualS 0 V.empty
  one  = DualS 1 V.empty
  DualS a da .+. DualS b db = DualS (a + b) (vzipPlus da db)
  DualS a da .*. DualS b db = DualS (a * b) (vzipPlus (vscale a db) (vscale b da))
\end{code}

Differentiating through WMC by computing in a semiring is not new: it is \emph{algebraic model counting}~\cite{kimmig2017}, and the dual-number (or \emph{gradient}) semiring was introduced for parameter learning in probabilistic logic programming~\cite{gutmann2008} and later generalized to arbitrary semirings~\cite{maene2025}.
In the same functional, BDD-based setting, Pluck~\cite{bowers2025} uses a dual-number semiring over its \texttt{rsdd}-based weighted model counting~\cite{pluck-ad}, as we do in Haskell.
The WMC implementation is polymorphic in the model weights, constrained only to the \hs{Semiring} type class (from \S\ref{sec:wmc}).
Hence, using dual numbers \hs{DualS} instead of the real numbers \hs{RealS}
is a one-line type change with no run-time dispatch, no configuration flags, and no separate code path.
\textbf{The four-line \hs{Semiring DualS} instance is the entire AD implementation.}

Formally, the reason why this is possible is that the WMC of a BDD is a \emph{multilinear polynomial} in the variable weights.
Each path from the root contributes a product of weights, and WMC sums these products.
Since WMC is expressed entirely in terms of semiring operations~\hs{(.+.)} and~\hs{(.*.)} applied to the weights,
the dual-number semiring computes both the polynomial's value and its gradient in a single forward pass.

As we saw from the final robot example, the complexity of finding the credal set representing an imprecise probabilistic program
grows exponentially with the number of Knightian choices.
Rather than enumerating all valuations of the Knightian choices, we can instead perform gradient descent to find the bounds of imprecise probabilities or expectations.

\paragraph{Seeding the partials.}
To optimize over the credal set, we seed each of the BDD nodes with a dual number.
The probabilistic variables are constant, so we seed them with zero gradient.
The Knightian weights are what we optimize over, so each one is initialized to $0.5$, with a partial of $-1$ in the low
weight and $+1$ in the high weight in its corresponding vector position.
Running WMC with these dual-number weights then yields both the expected score and gradient with respect to the Knightian variables in a single pass.

\paragraph{Learning optimal scores.}
With dual numbers in hand, we can perform standard gradient ascent (or descent) to find the best-case (or worst-case) expectation of a scoring function.
At each step, it computes the expected score and its gradient via WMC with dual numbers,
then adjusts in the direction of the gradient.
Since WMC is multilinear in the Knightian weights,
optimizing over a single Knightian variable with all others fixed is a linear program whose optimum lies at the extremes of $[0,1]$.
For multiple independent variables, the objective is multilinear over the product of simplices.
This is a non-convex problem in general, and while our optimizer is deterministic, random restarts are a standard remedy.
The same machinery can handle smooth functionals of the distribution (such as variance), where optima need not lie at extremal points.

\subsection{Single-Pass Conservative Inference via Interval Arithmetic} \label{sec:interval-wmc}

Another structure that supports semiring operations is probability intervals.
\begin{code}
data IntervalS = IntervalS !Double !Double

instance Semiring IntervalS where
  zero = IntervalS 0 0
  one  = IntervalS 1 1
  IntervalS a b .+. IntervalS c d = IntervalS (a + c) (b + d)
  IntervalS a b .*. IntervalS c d = IntervalS (a * c) (b * d)
\end{code}

Interval-valued WMC provides a complementary inference method:
a single-pass, linear-time sound outer approximation of the credal set bounds, trading tightness for speed.
This is useful as an efficient screening before committing to the more expensive enumeration or gradient methods.
The approximation is conservative but not tight, for two reasons.

Firstly, for interval values, Knightian variables $v$ are initialized with weights $w^\bot_v = w^\top_v = [0,1]$,
encoding that both the low branch and the high branch have unknown weight in $[0,1]$.
However, when these intervals are propagated through the algorithm, the knowledge that $w^\bot_v + w^\top_v = 1$ is lost.
This is the \emph{dependency problem} of interval arithmetic~\cite{Moore1966}.
The second limitation with interval-valued WMC is that each BDD guard is processed separately,
so the same Knightian variable can be evaluated differently,
contradicting the semantics and giving an over-approximation of bounds.

Even when the Knightian choices are not evaluated consistently, the result can still be useful.
It soundly encloses the credal set's true lower and upper probabilities, generally more loosely than exact enumeration but at a far lower cost.
The implementation again reuses the semiring abstraction, exposing the following methods.
\begin{code}
intervalProbabilityApprox :: Ord a => Imp g a -> (a -> Bool) -> (Double, Double)
intervalExpectationApprox :: Ord a => Imp g a -> (a -> Double) -> (Double, Double)
marginalApprox            :: Ord a => Imp g a -> [(a, Double, Double)]
\end{code}
For example, running interval-valued WMC on the \hs{dependent} example from \S\ref{sec:correlation}
decouples the two Knightian choices labelled \hs{"a1"}.
The resulting expectation bounds are thus resolved over the larger \hs{independent} credal set, rather than the \hs{dependent} one (see Figure~\ref{fig:credal-names}).
This still gives the correct marginals, but if we have a scoring function, \hs{score}, that assigns $0$ to~\hs{R}, $10$ to~\hs{G} and $-10$ to~\hs{B},
then running \hs{intervalExpectationApprox dependent score} does not track that~\hs{G} and~\hs{B}
are correlated, giving the wider bounds of $[-5, 5]$ rather than the precise $[0,0]$.

\subsection{Performance of Inference} \label{sec:performance}

The goal of this pearl is to demonstrate that a usable DSL can be idiomatically embedded into Haskell.
Although performance is not our emphasis, the choice of inference method matters as models grow, which we illustrate on a scalable family of programs.

We generalize the Ellsberg urn of \S\ref{sec:intro} to $N$ colours: colour~$0$ has known probability $1/N$, and the remaining $(N{-}1)/N$ mass is split over colours $1,\ldots,N{-}1$ in unknown proportion.
The split is decided by a form of \emph{stick-breaking}~(cf.~\cite{connor1969}) over $N{-}2$ Knightian coins: the first coin chooses between colour~$1$ and the remaining mass, the second between colour~$2$ and the rest, and so on, with colour~$N{-}1$ reached when no coin fires.
This is the \hs{ellsberg} program of \S\ref{sec:intro} (there $N=3$, with a single Knightian coin) extended to a chain of $N{-}2$ coins.

\begin{figure}
  \includegraphics[width=0.7\textwidth]{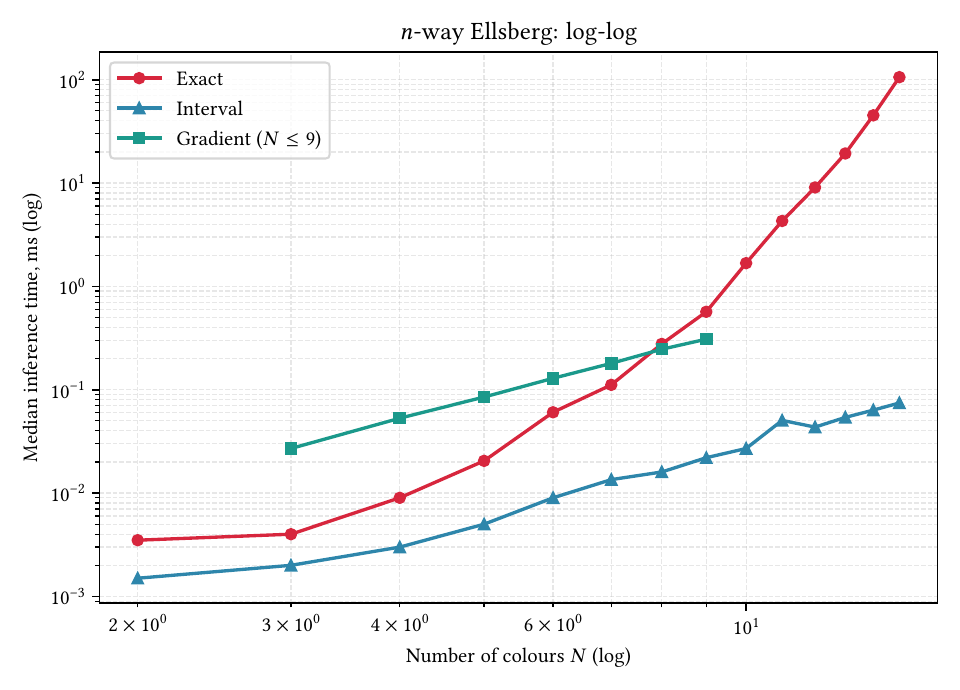}
  \caption{Inference time vs.\ number of colours $N$ on the $n$-way Ellsberg urn (log-log axes).
    Exact enumeration is exponential ($2^{N-2}$ Knightian valuations); interval WMC is polynomial.
    Gradient descent is polynomial (a fixed number of WMC passes per query, hence a larger constant); our current optimizer stays accurate to $N=9$.
    Medians of 10 runs, GHC~9.4.4 \texttt{-O2}.
  }
  \label{fig:benchmark}
  \Description{Log-log plot of inference time against the number of colours.}
\end{figure}

Figure~\ref{fig:benchmark} shows the time to bound $P(\mathsf{colour} = N{-}1)$ for $N = 2, \ldots, 15$.
The program compiles to a BDD of size $\mathcal{O}(N)$.
Naive exact enumeration of all $2^{N-2}$ Knightian valuations is exponential in~$N$; interval WMC is a single weighted model count over the BDD and is polynomial.
Gradient descent is likewise polynomial: each query runs a fixed number of ascent/descent steps, each a single weighted model count over the $\mathcal{O}(N)$ BDD, so its cost is $\mathcal{O}(N)$ with a larger constant than interval WMC.
On this family, our deterministic optimizer stays accurate to $N = 9$, so we plot it that far.

We anticipate that state-of-the-art knowledge-compilation techniques, such as variable order heuristics~\cite{Rudell1993, Panda1995} and sentential decision diagrams~\cite{darwiche2011sdd},
would lower the absolute constants further, but have no impact on the semantics or DSL interface.

\section{Further Examples}\label{sec:further-examples}

We close with two further examples that revisit features introduced above. The first, a Monty Hall variant, shows a Knightian choice that turns out to be vacuous, and the second, an imprecise two-child problem, illustrates conditioning.

\subsection{Monty Hall with Knightian Host Bias} \label{sec:montyhall}

In the Monty Hall problem, a contestant on a game show is standing in front of three doors.
A car is behind one door, while a goat is behind each of the others.
The player chooses a door, call it Door~1. The host, who knows which door the car is behind,
opens a different door, revealing a goat, and offers the player the chance to switch their choice.
In standard probability theory, switching to the other door wins with probability $2/3$.

A subtlety arises when the car is behind the door the player initially chose, Door~1.
The host must choose which of the two remaining doors to open, because behind both are goats.
In the original game, the choice is assumed uniform, but in reality, the host's bias is unknown.
We can model it as a Knightian choice.
\begin{code}
data Door = Door1 | Door2 | Door3

montyHall :: Imp '["host_bias"] Bool
montyHall = Imp.do
  c1       <- flip (1/3)
  c2       <- flip (1/2)
  hostBias <- knight @"host_bias"
  let car  = if c1 then Door1 else if c2 then Door2 else Door3
      host = case car of
               Door1 -> if hostBias then Door2 else Door3
               Door2 -> Door3
               Door3 -> Door2
      switchTo = case host of
                   Door2 -> Door3
                   Door3 -> Door2
  Imp.return (switchTo == car)
\end{code}
In this example, running \hs{intervalProbability montyHall id} yields the credal set $[2/3, 2/3]$.
Despite the Knightian uncertainty in the program, we get a precise answer.
Hence, we can conclude that the host bias doesn't impact the probability of the player winning if they switch.
The type signature \hs{Imp '["host\_bias"] Bool} records that the program has one degree of Knightian uncertainty, but it is impotent.

\subsection{The Imprecise Two-Child Problem} \label{sec:twochild}

A common paradox in conditional probability is the \emph{two-child problem}:
a family has two children, at least one of which we know is a boy.
Classic probability tells us that since the outcome of two daughters is impossible,
the chance that both are boys is $1/3$.

Now consider that the family has adopted one of the children from an interstellar exchange programme.
Human birth statistics are well-known, but we cannot assign a prior to the alien child's status.
Knowing at least one is a boy, we may ask for the probability that both are.
\begin{code}
twoChild :: Imp '["alien"] Bool
twoChild = Imp.do
  humanBoy <- flip 0.5
  alienBoy <- knight @"alien"
  observe (humanBoy || alienBoy)
  Imp.return (humanBoy && alienBoy)
\end{code}
The \hs{observe} statement is the DSL's conditioning operator,
restricting subsequent inference to executions in which the predicate holds.

Running \hs{intervalProbability twoChild id} yields the credal set $[0,1/2]$, which contains the classical answer.
The extreme points have natural interpretations.
When the alien resolves to a boy, the observation is automatically satisfied
and provides no information about the human, so the probability is $1/2$.
When the alien resolves to a girl, the chance that both are boys must be $0$.

\section{Concluding Remarks and Related Work} \label{sec:related}

\paragraph{Probabilistic programming and knowledge compilation.}
Probabilistic programming has a long history in functional languages (e.g.~\cite{erwig2006}).
Modern Haskell probabilistic programming languages, such as monad-bayes~\cite{scibior2018}, target continuous
distributions with sampling-based inference;
our work instead targets discrete models with exact inference
and adds imprecise probability.
For discrete models, binary decision diagrams (BDDs)~\cite{bryant1986} are a well-studied compilation target for exact inference
via weighted model counting (WMC)~\cite{darwiche2002,darwiche2009,chavira2008}.
ProbLog~\cite{deraedt2007} pioneered this approach in probabilistic logic programming,
and Kimmig et al.~\cite{kimmig2017} generalized WMC to an algebraic setting
parametrized by a semiring, which we exploit in \S\ref{sec:ad} and \S\ref{sec:interval-wmc}.
Dice~\cite{holtzen2020} compiles discrete probabilistic programs to BDDs
for exact inference, using modular compilation to scale to large models.
Pluck~\cite{bowers2025} extends this with lazy knowledge compilation
to support higher-order functions and recursion.
The underlying Rust library \texttt{rsdd} provides semiring-parametrized WMC,
which Pluck instantiates with a dual-number semiring for gradients~\cite{pluck-ad}.
The primary difference between our system and these BDD-based languages is
our treatment of \emph{imprecise probability}:
our programs contain both probabilistic and Knightian uncertainty,
and inference computes interval-valued bounds rather than point probabilities.
The use of standard and plain BDD and WMC machinery is key.
Imprecise probability slots into the existing discrete probabilistic programming pipeline with no new compilation infrastructure.
The only additions are the graded-monad layer that tracks Knightian names at the type level, and a single semiring-parametric weighted model counting that serves several inference methods.

McIver and Morgan~\cite{mciver2005} study demonic non-determinism alongside probabilistic choice in a program refinement setting.
Our worst-case credal optimization (\S\ref{sec:ad}) is analogous to their demonic choice,
but our motivation is modelling of epistemic uncertainty rather than program verification.

\paragraph{Graded monads and effect tracking.}
The type-level tracking that distinguishes our system
draws on the graded monad for imprecise probability,
$T_g(a) = (2^g \to D(a))$~\cite{liellcock2025},
which assigns a distribution over~$a$ for each valuation of the finite set
of Knightian names~$g$.
Graded monads were introduced by Katsumata~\cite{katsumata2014} and applied
to effect tracking by Orchard, Liepelt, and Eades~\cite{orchard2019} in the Granule language;
Orchard and Petricek~\cite{orchard2014} explored encoding such systems
in Haskell using type-level machinery, an approach we follow here.
An alternative design would use algebraic effect handlers, but the monoidal grading that tracks and merges Knightian names
does not currently arise in the handler-based setting,
where effects are characterized by operations and interpretations
rather than by a compile-time grade.
Our monad \hs{Imp} is graded by finite sets of names with union as the monoid operation,
the grade identified in~\cite{liellcock2025} for modelling imprecise probability.
The crucial benefit of this grading is \emph{commutativity}:
the convex powerset monad (the standard monad for sets of distributions)
is not commutative, so the order of composition can affect the credal set;
the graded monad of~\cite{liellcock2025}, studied string-diagrammatically
by Sarkis and Zanasi~\cite{sarkis2025}, restores commutativity when name sets are disjoint,
and our \hs{Merge} type family enforces this precondition at compile time.

Lew et al.~\cite{lew2020} introduced trace types for probabilistic programs,
tracking random sampling in the type system via a graded monad in Haskell.
Their work inspired our type-level tracking of the grade.
However, the grades track different information:
Lew et al.'s trace types record which random choices a program makes,
with the key property being measurability,
whereas our grades record the Knightian choices,
with the key property being disjointness (ensuring that names from independent
sub-computations do not collide).
This paper shows that Haskell's type-level lists, GADTs, and \hs{QualifiedDo}
provide a practical encoding of the graded monad from~\cite{liellcock2025}.

\paragraph{Future directions.}
A fourth possible semiring instantiation is convex polyhedra over the Knightian variables, computing the exact credal set in a single pass.
The trade-off is representation size: such polyhedra are multilinear polynomials whose monomial count can grow exponentially in the number of Knightian variables,
and optimizing over them is NP-hard in general, though BDD structure-sharing should mitigate this in practice.

Our simple prototype uses BDDs, which suffice for the examples in this paper
but are sensitive to variable ordering~\cite{bryant1986}.
Sentential decision diagrams~\cite{darwiche2011sdd},
as used by the \texttt{rsdd} library underlying Dice~\cite{holtzen2020}
and Pluck~\cite{bowers2025},
would offer greater robustness and efficient bottom-up compilation.
Extending the inference engine to continuous distributions would be a further
challenge, although first attempts could build on other techniques (e.g.~\cite{holtzen2024})
or alternative approaches from the MDP literature.

\paragraph{Summary.}
We have presented a functional pearl showing that Haskell's type system
provides a natural home for imprecise probabilistic programming.
Probabilistic and Knightian coin flips both compile to BDD variables.
The same semiring-polymorphic WMC code serves three different inference methods, and the type-level names ensure that independently specified uncertainties
cannot accidentally interfere.
The result is a small library\footnote{Available at: \url{https://github.com/jacklc3/imp}} (approximately 1,200 lines of Haskell)
bridging the mathematical theory of imprecise probability~\cite{liellcock2025}
and the practical tools of discrete probabilistic programming,
aimed at robust decision-making under model uncertainty.

\begin{acks}
  We particularly thank Alex Lew, whose trace types~\cite{lew2020} inspired our type-level approach,
  and who pointed us to the gradient support in Pluck~\cite{pluck-ad} that underlies our
  gradient-based inference over credal sets.
  We also thank Todd Millstein, the ARIA Safeguarded AI community, and the Oxford group for helpful discussions.
  This research was supported by the Clarendon Scholarship, the ERC Consolidator Grant BLAST, AFOSR Project FA9550-21-1-0038, and the ARIA programme on Safeguarded AI.
\end{acks}

\bibliographystyle{ACM-Reference-Format}
\bibliography{pearl}

\end{document}